# The Quantified Body: Identity, Empowerment, and Control in Smart Wearables

Maijunxian Wang

University of California, Berkeley

mjxwang@berkeley.edu

## Introduction

In the age of ubiquitous computing, the body has become not only visible but also measurable, analyzable, and programmable. Smart wearables—devices like the Apple Watch, Fitbit, and Oura Ring—offer users real-time insights into their physical and emotional states. Marketed as tools of self-optimization and preventative care, they form a new interface between individuals and digital infrastructures: the "quantified self," made wearable (Lupton, 2016).

Yet beneath this promise lies a more complex system. These devices do not merely display metrics—they collect, transmit, and algorithmically process streams of biometric data. This data often enters systems not just designed to inform but to predict, steer, and monetize behavior. Who owns this data? Where does it go? What happens when our heartbeat, sleep cycle, or step count becomes an input for risk prediction models used by insurers, employers, or platforms?

While wearables offer an unprecedented sense of control, they also embed users into architectures of surveillance and value extraction. In other words, the empowering interface comes with invisible strings.

To unpack these tensions, this paper draws on three critical theoretical frameworks: Gilles Deleuze's (1992) concept of the *control society*, Shoshana Zuboff's (2019) theory of *surveillance capitalism*, and Nick Couldry and Ulises Mejias's (2019) framework of *data colonialism*. These theories reveal how seemingly personal and helpful technologies may encode deeper systems of power and control.

Focusing on smart wearables as a case study, I argue that these devices have evolved from self-tracking tools into predictive-governance infrastructures. They translate bodily rhythms into actuarial risk scores, reinforcing what is called the *post-consent regime* (Lindner, 2020)—a system in which users technically grant permission, but structural autonomy is eroded by default design.

Throughout this paper, I explain and apply these concepts to examine how wearables reshape not only health behavior, but also the lived experience of embodiment, identity, and participation in digital society. While existing literature explores privacy and behavioral nudging, fewer studies interrogate how biometric feedback loops reconfigure bodily autonomy into quantifiable, governable, and extractable forms.

## Theoretical Foundations: From Control Societies to Data Colonialism

Understanding smart wearables requires moving beyond their function as health tools and instead consider them as embedded mechanisms of governance. In the following section, I utilize three theoretical frameworks—control society, surveillance capitalism, and data colonialism—to illuminate how these devices operate through feedback, extraction, and codification, enrolling the body into new regimes of power. This analysis offers a deeper understanding of the relationship between bodily data, technological interfaces, and political economy.

Gilles Deleuze's (1992) notion of the *control society* describes a shift from Foucault's disciplinary institutions to continuous, flexible forms of modulation. In this paradigm, control no longer confines—it flows. As Deleuze writes, "you do not confine people with a highway, but by making it endlessly traversable." Smart wearables embody this shift. Rather than impose external commands, they induce self-regulation through constant feedback. Users track themselves, interpret metrics, and adapt behavior accordingly—governed not by force, but by feedback loops.

Shoshana Zuboff's (2019) concept of *surveillance capitalism* explains the economic logic behind this control. For Zuboff, behavioral data has become a raw material, appropriated and monetized by corporations. Wearables, in this view, are not just health aids but data-harvesting instruments. They feed user inputs into predictive models used by employers, insurers, and marketers. What appears as personal empowerment is, structurally, a form of asymmetrical information extraction—what Zuboff calls "instrumentarian power."

Couldry and Mejias (2019) extend this critique through their theory of *data colonialism*. They argue that modern data systems replicate colonial dynamics—appropriating personal life as a resource. Just as empires extracted land and labor, today's infrastructures extract experience, turning our bodies into perpetual sources of value. In this view, wearables don't simply "track" bodies—they transform lived embodiment into monetizable flow, often without meaningful consent.

Taken together, these theoretical frameworks reveal that wearables are not neutral technological interfaces but nodes within broader systems of governance. They help us understand how convenience can obscure coercion, how optimization can mask normalization, and how seamless design can conceal structural asymmetries. In the sections that follow, these theories will guide our analysis of how smart wearables shape bodily autonomy, social identity, and everyday life—allowing us to see beyond the device and into the political economies it serves.

Yet to truly understand how governance operates through these devices, we must begin with their seductive appeal. The power of wearable technologies often starts not with coercion, but with the promise of health, autonomy, and efficiency. Control takes root precisely through empowerment. This creates a paradox: the very features that foster autonomy also deepen behavioral regulation. To grasp how this paradox unfolds, let us turn to the persuasive promise of empowerment and examine how it governs in the name of motivation.

## Empowerment: The Promise of Personalized Health and Agency

As previously discussed, these devices have become predictive governance tools—not only recording data, but actively shaping behavior through seemingly empowering features. To grasp why predictive governance is so persuasive, we must first understand the real, embodied benefits that seduce users into self-discipline. Their global popularity is evident: according to International Data Corporation (2024), the global wearables market shipped approximately 538 million devices in 2024, with projected growth of around 6% in 2025. This growth stems from real benefits in health awareness, goal-setting, and preventative care. They bring the "quantified self" ideal into everyday life, helping users visualize and optimize their behaviors.

Wearables offer real-time feedback on metrics like step count, heart rate, and sleep patterns. This data transforms invisible bodily processes into actionable information, turning health management into a daily, self-directed practice. Studies show even simple features—like reminders to stand or daily step goals—can increase activity, especially among sedentary populations. According to Piwek et al. (2016), wearables are most effective when combined with persuasive design: gamified goals, social comparison, and haptic prompts. Users often report greater self-discipline and motivation, with the device acting as a personal coach. Cadmus-Bertram et al. (2015) conducted a randomized controlled trial demonstrating that a Fitbit-based intervention significantly increased daily step counts among sedentary adults, supporting the persuasive potential of self-tracking technologies.

Wearables also expand access to health data. By decentralizing monitoring, they offer value in regions with limited medical infrastructure. For people with chronic conditions, devices can track symptoms over time, enabling more informed interactions with healthcare providers and early warnings for conditions like atrial fibrillation. Beyond physical metrics, many wearables now track stress, recovery, and readiness—offering insight into mental health and well-being. Devices like the Oura Ring and WHOOP Band validate rest and balance, encouraging users to prioritize recovery alongside performance.

In professional and athletic contexts, wearables are used to manage exertion, optimize training, and prevent injury. In workplace wellness programs, some employees voluntarily use wearables to qualify for insurance incentives or set personal goals—gaining structure without third-party oversight.

However, this sense of empowerment is not equally accessible to all. It presumes digital literacy, stable internet connectivity, and discretionary income. More critically, the discourse of self-optimization aligns seamlessly with neoliberal values of discipline, efficiency, and productivity. The "empowered" subject is expected to become not only healthier, but also more compliant, efficient, and data-conscious.

Nonetheless, for many users, these tools offer genuinely motivating experiences. The structure and rhythm they provide resonate with desires for order, self-control, and measurable progress. This sense of "agency within guidance" often brings reassurance and enthusiasm, effectively masking the underlying mechanisms of behavioral governance. Precisely because of this, we must attend to what is sacrificed in exchange for agency: What is relinquished when care becomes computation, and motivation becomes metric? What kind of subject is being reconfigured in this shift?

These questions point to a central contradiction: the very features that claim to enhance autonomy are also key mechanisms of behavioral control and surveillance. The next section turns to this central contradiction, examining how technical design constructs empowerment and control not as opposites, but as mutually reinforcing forces.

## Control: Surveillance and the Biometric Assemblage

The original promise of empowerment has gradually evolved into a system of governance—one in which health-oriented features are leveraged to enforce actuarial compliance. Personalized feedback, step goals, and readiness scores serve not only to motivate users but also to align their behaviors with institutional norms. These "benefits" in turn become levers for pushing self-management toward institutional conformity, embedding the body within opaque predictive markets. While wearables promise autonomy, they operate within infrastructures designed to extract, process, and monetize biometric data. As these devices transition from lifestyle gadgets to formal tools in medical and employment settings, the boundary between self-tracking and systemic control becomes increasingly blurred.

At the heart of this control structure are continuous streams of biometric data—heart-rate variability, sleep cycles, GPS traces, and movement patterns. These data are routinely uploaded to proprietary servers, where they are used to refine algorithms, train predictive models, and are often sold to third parties such as employers or insurance providers. Recent legal analyses have revealed that many wearable devices transmit biometric data to external servers without obtaining meaningful informed consent. Although users technically "agree" to data sharing, this consent is often buried in lengthy privacy policies, default settings, or user interfaces designed for ease rather than transparency. As Mone and Shakhlo (2023) note, this creates a situation where compliance is achieved in form but not in substance—undermining the intent of data protection laws even when their formal requirements are met. The European Parliamentary Research Service (2025) similarly warns that although the GDPR guarantees a fundamental right to data protection, its enforcement has struggled to keep pace with the rapid proliferation of biometric technologies like wearables.

This architecture epitomizes the *surveillant assemblage* described by Haggerty and Ericson (2003): a decentralized network that recombines data fragments to build actionable profiles across time and context. Such flows recast the body as a continuously evaluated object; step goals, recovery scores, and heart-rate zones assign moral valence to performance, ranking and guiding users through predictive models.

In institutional contexts, this dynamic becomes even more pronounced. Employers incorporate wearable devices into wellness incentive programs, offering financial rewards or insurance discounts to encourage participation—nominally voluntary, yet socially coercive. At the same time, insurance companies increasingly exchange premium discounts for user data, shifting the logic of risk assessment from population-level statistics to individualized biometric markers. The compliant, data-sharing individual becomes the ideal consumer. Although users are technically allowed to opt out of these tracking mechanisms—by disabling certain features or choosing open-source devices—in practice, doing so often entails penalties: forfeiting insurance benefits, being excluded from employer programs, or being perceived as uncooperative or "noncompliant." As Aldana (2025) points out, some companies tie wellness incentives directly to wearable data, creating implicit penalties for employees who choose to retain their privacy. For example, merely refusing to share personal data can result in the loss of monthly health insurance discounts. Such mechanisms reinforce a system that appears voluntary on the surface but is coercive in practice, making opting out a costly and discouraging choice.

Thus emerges a troubling contradiction: wearables offer visibility and insight even as they embed bodies within circuits of control that are hard to perceive or escape. The quantified self becomes a predictable self—measurable, optimizable, and perpetually extractable. To resist this trajectory, we must ask whether current designs and policies serve autonomy or exploitation, and look instead to alternate design histories and emerging efforts that envision wearable futures grounded in care, transparency, and agency.

## Alternate Design Futures: From Historical Paths to Anti-Extractive Practices

Historical counter-examples reveal that predictive governance is not a technical inevitability, but a political design choice. The current convergence of wearable technologies and surveillance capitalism is one trajectory among many. Alternative paths have long existed—and continue to emerge—that center care, dignity, and autonomy over optimization and control.

One compelling example comes from the work of designer and metalsmith Mary Ann Scherr, as documented by Kayleigh Perkov (2022). In the 1960s and 70s, Scherr crafted wearable medical devices that prioritized embodied presence and human relationships. Her designs integrated heart monitors and respiratory trackers into jewelry—necklaces, pendants, and bracelets—that were functional, beautiful, and deeply personalized. These devices were often co-developed with patients and doctors, grounding technological monitoring in collaboration rather than institutional surveillance. As Perkov notes, this philosophy of design emphasized care over compliance, and presence over performance.

The aesthetics of these devices were far from neutral. Unlike today's fitness trackers—designed to be sleek, standardized, and to "blend in" seamlessly—Scherr's creations made technology visible, tangible, and emotionally resonant. This deliberate visibility was not merely an artistic choice but an ethical stance. By making the body's condition perceptible to others, wearables ceased to be private surveillance tools and instead became shared, relational instruments of care. In this sense, bodily presentation itself became a medium of technological ethics, exposing the power relations embedded in design. Where contemporary devices often conceal their monitoring functions under the guise of "natural" or "invisible" design, Scherr's work foregrounded transparency, intimacy, and reflection.

Scherr's legacy invites a broader reflection on the moral dimensions of wearable design. This legacy invites us to rethink what wearables could be: not just interfaces for data extraction, but sociotechnical artifacts embedded in moral and affective economies. While today's dominant design values emphasize scale, automation, and seamlessness, other values—such as transparency, agency, relationality, and narrative—can guide alternative futures.

Contemporary anti-extractive design practices illustrate this shift. Projects like OpenBCI, an open-source brain-computer interface platform, enable users to monitor and engage with their neurophysiological data on their own terms. Grassroots DIY health monitor initiatives further protect user agency by keeping data local and encrypted, fostering community-based health data trusts. These designs shift emphasis from corporate metrics to communal autonomy.

Beyond grassroots innovation, even mainstream design practices are beginning to show signs of resistance to extractive logic. Examples include on-device processing (where biometric data is analyzed locally), opt-in data sharing, and haptic consent prompts (such as a vibration before transmission), which introduce friction into an otherwise frictionless data landscape. Though not revolutionary alone, they reflect a more participatory and transparent design ethos.

This opens up a further possibility: shifting the purpose of wearables from individual self-optimization to collective well-being. Speculative concepts like environmental alert necklaces that flash when air quality deteriorates, or community stress monitors that aggregate anonymous biometric data, shift focus from personal optimization to public care. Such designs align wearable technology with social solidarity rather than self-surveillance.

Finally, the growing interest in "design justice" reinforces this alternative vision. Scholars and activists like Sasha Costanza-Chock (2020) argue that technologies must be co-designed with the communities most affected by surveillance and inequality. *In this view, wearable design is a political practice—reinforcing or resisting structural harms.*

In sum, wearables are not inherently extractive. Their current architecture reflects economic and political decisions—not technical necessities. By drawing on historical precedents and contemporary innovations, we can imagine a future where wearables support not just longer lives, but more livable ones—where care and agency over data-driven discipline.

Achieving this future, however, requires more than visionary prototypes; it demands enforceable governance structures capable of scaling these values. Without corresponding changes in regulation and data infrastructure, even the most ethically designed devices risk being subsumed into extractive systems. Design justice alone cannot counter surveillance capitalism—it must be matched by structural reforms in how wearable data is governed.

## Policy and Governance: From Consent to Justice

As smart wearables increasingly shape how our bodies are monitored, evaluated, and acted upon, they participate in what scholars describe as a "post-consent regime"—a system where users technically agree to data collection, but true autonomy is undermined by opaque defaults and structural asymmetries. In this context, data-justice governance must attack the structure, not the checkbox. Regulations such as the European Union's General Data Protection Regulation (GDPR), the California Consumer Privacy Act (CCPA), and the upcoming EU Artificial Intelligence Act offer frameworks for protecting personal data, enforcing consent, and defining ethical standards for data-driven technologies.

Yet in practice, these legal instruments have done little to disrupt the logic of extraction that governs the wearable tech ecosystem.

One key limitation lies in the overreliance on individual consent as the cornerstone of data governance. While users are technically required to agree to data collection, the structures of this consent are deeply flawed. As numerous scholars have argued, the notion of *informed consent* becomes hollow when users are faced with opaque privacy policies, complex interfaces, or binary "agree/decline" options that condition access to functionality. This is especially true for wearables, where the device's value is deeply tied to seamless integration and constant connectivity. In this context, "frictionless design" undermines genuine autonomy: the easier it is to say yes, the harder it is to understand what one is agreeing to.

The shortcomings of consent reveal a deeper issue: treating data as a private asset subject to contractual transfer overlooks the structural conditions under which that data is produced, interpreted, and acted upon. This is where the concept of *data justice* becomes crucial. Instead of asking whether users have *agreed*, data justice asks whether the entire system of data relations—who collects, who benefits, who is harmed—is equitable, transparent, and democratically accountable. It shifts the focus from individual transactions to collective rights.

A data justice approach to wearable governance calls for a transformation in both technological architecture and institutional oversight. While these proposals may appear ambitious, they draw on existing principles from privacy engineering, AI ethics, open data standards, and emerging experiments in democratic data stewardship. Applied to wearables, these principles translate into five key intervention pathways that challenge extractive norms and center transparency, equity, and user agency.

1.On-Device Computation

Wearable data should be processed locally, minimizing exposure by avoiding remote uploads. This reduces dependency on centralized servers and limits commodification, even if technically challenging.

2.Data Portability & Interoperability

Users must be able to export biometric data in open formats and transfer it across platforms. Such interoperability restores user control and supports audits, research, and decentralized development.

3.Community Data Trusts

Instead of relying on individual consent, wearable data could be managed by data trusts—independent entities that negotiate collective terms and enforce accountability, especially in workplaces or high-risk sectors.

4.Active, Embodied Consent

Interfaces should support embodied consent, using tactile signals (e.g., vibration or LED flash) before transmitting data. These prompts reintroduce awareness and challenge the passive flow of information.

5.Algorithmic Accountability

As wearables generate risk scores and recommendations, algorithms must be transparent and auditable. Public registries, technical standards, and the right to contest automated decisions are essential safeguards.

Taken together, these interventions constitute a shift in paradigm: from *permission-based regulation* to *structure-oriented justice*. They recognize that the harms of wearable technologies are not just about privacy breaches, but about the normalization of extractive logic, the erosion of bodily sovereignty, and the enclosure of future possibilities. As Couldry and Mejias (2019) argue, data relations must be reimagined not merely as economic exchanges but as political relationships, shaped by power, history, and inequality.

Importantly, none of these changes can be accomplished by technology alone. Institutional enforcement, public awareness, and democratic participation are essential. Just as we regulate pharmaceuticals and food production, so too must we regulate the infrastructures that increasingly govern our bodies and identities. Smart wearables are not neutral gadgets; they are governance tools in disguise.

If the quantified body is here to stay, then the politics of its design, data, and regulation will shape not just how we live—but who we are allowed to become.

### Reclaiming the Quantified Body: From Compliance to Collective Control

Smart wearables, on the surface, promote individual self-optimization—but beneath this promise lies a new mode of governance: prediction in name, compliance in practice. Once the body is transformed into a stream of data, it ceases to be merely a biological entity and becomes an object of algorithmic evaluation, commercial manipulation, and institutional discipline.

As Deleuze's theory of control society, Zuboff's critique of surveillance capitalism, and Couldry and Mejias's concept of data colonialism all make clear, technology is not neutral. The power structures embedded in its design determine whom it serves—and whom it sacrifices. Today's wearables are not medically neutral tools; they are extensions of capitalist governance logic, institutionalizing the commodification of bodies, the predictability of labor, and the individualization of risk.

Yet both history and the present remind us that another technological future is possible. From Mary Ann Scherr's human-centered designs to anti-extractive practices in open-source communities, wearable technologies can be reimagined as instruments of care, solidarity, and democratic engagement. This is not merely a matter of design—it is a matter of political stance.

True transformation must begin at the level of data governance, shifting from a paradigm of individual consent to one of collective control. From "user" to "citizen," from "product" to "right," the future of wearables must be built upon institutional accountability, public oversight, and shared decision-making.

The quantified body must not be one that is predicted, managed, and dispossessed. It must be a body that can speak, that holds rights, and that participates in collective governance. This is the central question of technological politics: what kind of body do we want, and in what kind of society do we want it to live?

Data may predict behavior, but it cannot predict awakening. Only collective action can end this invisible expropriation. The choice is still ours—but not as users. It is as citizens that we must raise our voices—only then will technology belong to us, rather than control us.

## References

Ajana B. (2017). Digital health and the biopolitics of the Quantified Self. *DIGITAL HEALTH*. 2017;3. doi:10.1177/2055207616689509

Aldana, S. (2025, February 28). Common mistakes corporate health programs make with wearable devices. *WellSteps*. https://www.wellsteps.com/blog/2020/01/02/corporate-health-programs/

Cadmus-Bertram, L., Marcus, B. H., Patterson, R. E., Parker, B. A., & Morey, B. L. (2015). Use of a Fitbit device as part of a physical activity intervention for obese, postmenopausal women: A randomized controlled trial. *JMIR mHealth and uHealth*, *3*(4), e2145. https://doi.org/10.2196/mhealth.4229

Costanza-Chock, S. (2020). *Design justice: Community-led practices to build the worlds we need*. MIT Press.

Couldry, N., & Mejias, U. A. (2019). Data Colonialism: Rethinking Big Data's Relation to the Contemporary Subject. *Television & New Media*, *20*(4), 336-349. https://doi.org/10.1177/1527476418796632

Deleuze, G. (1992). Postscript on the Societies of Control. *October*, *59*, 3–7. http://www.jstor.org/stable/778828

European Parliamentary Research Service. (2025, January 28). Understanding EU data protection policy [Briefing]. *European Parliament Think Tank*. https://epthinktank.eu/2025/01/28/understanding-eu-data-protection-policy-3/

Haggerty, K. D., & Ericson, R. V. (2003). The surveillant assemblage. *The British Journal of Sociology*, *51*(4), 605–622. https://doi.org/10.1080/00071310020015280

International Data Corporation. (2024, October 10). IDC forecasts continued growth for wearables but growth will be slower [Press release]. https://www.idc.com/getdoc.jsp?containerId=prUS52615024

Kayleigh, C. P. (2022). Space suits and gas masks: Mary Ann Scherr and an alternative view of personal technology. *Journal of Design History*, *35*(4), 401–417. https://doi.org/10.1093/jdh/epab051

Lindner, P. (2020). Molecular politics, wearables, and the aretaic shift in biopolitical governance. *Theory, Culture & Society*, *37*(3), 71–96. https://doi.org/10.1177/0263276419894053

Lupton, D. (2016). Self-tracking, health and medicine. *Health Sociology Review*, *26*(1), 1–5. https://doi.org/10.1080/14461242.2016.1228149

Mone, V., & Shakhlo, F. (2023). Health data on the go: Navigating privacy concerns with wearable technologies. *Legal Information Management*, *23*(3), 179–188. https://doi.org/10.1017/S1472669623000427

Piwek, L., Ellis, D. A., Andrews, S., & Joinson, A. (2016). The rise of consumer health wearables: Promises and barriers. *PLOS Medicine, 13*(2), e1001953. https://doi.org/10.1371/journal.pmed.1001953

Whitson, J. R. (2013). Gaming the quantified self. *Surveillance & Society*, *11*(1/2), 163–176. https://doi.org/10.24908/ss.v11i1/2.4454

Zuboff, S. (2019). *The age of surveillance capitalism: The fight for a human future at the new frontier of power*. PublicAffairs.